\documentclass[12pt,onecolumn]{revtex4}
\begin{document}

\title{Thermodynamics of charged rotating black branes in
 Brans-Dicke theory with quadratic scalar field potential}
\author{M. H. Dehghani$^{1,2}$ \footnote{email address:
mhd@shirazu.ac.ir}, J. Pakravan$^{1}$ and S. H. Hendi$^{1}$}
\affiliation{$^1$Physics Department and Biruni Observatory,
College of Sciences, Shiraz
University, Shiraz 71454, Iran\\
$^2$Research Institute for Astrophysics and Astronomy of Maragha
(RIAAM), Maragha, Iran}
\begin{abstract}
We construct a class of charged rotating solutions in
$(n+1)$-dimensional Maxwell-Brans-Dicke theory with flat horizon
in the presence of a quadratic potential and investigate their
properties. These solutions are neither asymptotically flat nor
(anti)-de Sitter. We find that these solutions can present black
brane, with inner and outer event horizons, an extreme black brane
or a naked singularity provided the parameters of the solutions
are chosen suitably. We compute the finite Euclidean action
through the use of counterterm method, and obtain the conserved
and thermodynamic quantities by using the relation between the
action and free energy in grand-canonical ensemble. We find that
these quantities satisfy the first law of thermodynamics, and the
entropy does not follow the area law.
\end{abstract}

\maketitle

\section{Introduction\label{Intro}}

There has been much more interest in alternative theories of
gravity in recent years. This is due to the fact that at the
present epoch the Universe expands with acceleration instead of
deceleration along the scheme of standard Friedmann models
\cite{Riess}. One of the alternative theory of gravity is the
scalar-tensor gravity pioneered by Jordan, Brans and Dicke (JBD)
\cite{Bran}. In recent years this theory has attracted a great
deal of attention, in particular in the strong field domains, as
it arises naturally as the low energy limit of many theories of
quantum gravity such as the supersymmetric string theory or the
Kaluza-Klein theory. It has been shown that the JBD theory seems
to be better than the Einstein gravity for solving the graceful
exit problem in the inflation model \cite{La}. This is because the
scalar field in the BD theory provided a natural termination of
the inflationary era via bubble nucleation without the need for
finely tuned
cosmological parameters. This theory contains an adjustable parameter $%
\omega $ that represents the strength of coupling between scalar
field and the matter.

Due to highly non-linear character of JBD theory, a desirable
pre-requisite for studying strong field situation is to have
knowledge of exact explicit solutions of the field equations. Four
forms of static spherically symmetric vacuum solution of the BD
theory in four dimensions are available in the literature which
are constructed by Brans himself \cite{Bran2}. However, it has
been shown that among these four classes of the static spherically
symmetric solutions of the vacuum Brans-Dicke theory of gravity
only two are really independent \cite{Bha1}, and only one of them
is permitted for all values of $\omega $. Although this class of
solutions, in general, gives rise to naked singularity, for some
particular choices of the solution's parameters it represents a
black hole different from Schwarzschild one \cite {Cam}. The other
class of solutions is valid only for $\omega <-3/2$ which implies
non-positive contribution of matter to effective gravitational
constant and thus a violation of weak energy condition
\cite{Bran2}. Static charged solutions of Brans-Dicke-Maxwell
gravity have been investigated in \cite{Cai}, and the non-trivial
Kerr-Newman type black hole solutions different from general
relativistic solutions have been constructed in JBD for
$-5/2<\omega <-3/2$ \cite{Kim}. Black hole solutions with
minimally and conformally coupled self-interacting potential have
been found in three \cite {Zan1} and four \cite{Zan2} dimensions
in the presence of cosmological constant. Constructing new exact
solutions of JBD theory from the known solution has been also
considered in \cite{DSh}. Till now, charged rotating black hole
solutions for an arbitrary value of $\omega $ has not been
constructed. In this paper, we want to construct exact charged
rotating black hole solutions in Brans-Dicke theory for an
arbitrary value of $\omega $ and investigate their properties.

The outline of our paper is as follows. In Sec. \ref{Con}, we give
a brief review of the field equations of Brans-Dicke theory in
Jordan (or string) and Einstein frames. In Sec. \ref{Sol}, we
obtain charge rotating solution in $(n+1)$-dimensions with $k$
rotation parameters. In Sec. \ref{Therm}, we obtain the finite
action, and compute the conserved and thermodynamic quantities of
the $(n+1)$-dimensional black brane solutions with a complete set
of rotational parameters. We also show that these quantities
satisfy the first law of thermodynamics. We finish our paper with
some concluding remarks.

\section{Field Equation and Conformal Transformation\label{Con}}

Long-range forces are known to be transmitted by the tensor
gravitational field $g_{\mu \nu }$ and the vector electromagnetic
field $A_{\mu }$. It is natural then to suspect that other
long-range forces may be produced by scalar fields. Such theories
have been suggested since before relativity. The simplest theory
in which a scalar field shares the stage with gravitation is that
of Brans-Dicke theory. In $n$ dimensions, the action of the
Brans-Dicke-Maxwell theory with one scalar field $\Phi $ and a
self-interacting potential $V(\Phi )$ can be written as

\begin{equation}
I_{G}=-\frac{1}{16\pi }\int_{\mathcal{M}}d^{n+1}x\sqrt{-g}\left(
\Phi \mathcal{R}\text{ }-\frac{\omega }{\Phi }(\nabla \Phi
)^{2}-V(\Phi )-F_{\mu \nu }F^{\mu \nu }\right) ,  \label{I1}
\end{equation}
where $\mathcal{R}$ is the Ricci scalar, $F_{\mu \nu }=\partial
_{\mu
}A_{\nu }-\partial _{\nu }A_{\mu }$ is the electromagnetic tensor field, $%
A_{\mu }$ is the vector potential, $\omega $ is the coupling
constant, $\Phi $ denotes the BD scalar field and $V(\Phi )$ is a
self-interacting potential for $\Phi $. Varying the action
(\ref{I1}) with respect to the metric, scalar and vector fields
give the field equations as

\begin{eqnarray}
&& G_{\mu \nu } =\frac{\omega }{\Phi ^{2}}\left( \nabla _{\mu
}\Phi
\nabla_{\nu }\Phi -\frac{1}{2}g_{\mu \nu }(\nabla \Phi )^{2}\right) -\frac{%
V(\Phi )}{2\Phi }g_{\mu \nu }+\frac{1}{\Phi }\left( \nabla _{\mu
}\nabla
_{\nu }\Phi -g_{\mu \nu }\nabla ^{2}\Phi \right)  \nonumber \\
&&\hspace{1cm}+\frac{2}{\Phi }\left( F_{\mu \lambda }F_{\nu }^{\text{ }%
\lambda }-\frac{1}{4}F_{\rho \sigma }F^{\rho \sigma }g_{\mu \nu
}\right) ,
\label{fil1} \\
&& \nabla ^{2}\Phi =-\frac{n-3}{2\left[ \left( n-1\right) \omega +n\right] }%
F^{2}+\frac{1}{2\left[ \left( n-1\right) \omega +n)\right] }\left[
(n-1)\Phi
\frac{dV(\Phi )}{d\Phi }-\left( n+1\right) V(\Phi )\right] ,  \label{fil2} \\
&& \nabla _{\mu }F^{\mu \nu }=0,  \label{fil3}
\end{eqnarray}
where $G_{\mu \nu }$ and $\nabla _{\mu }$ are the Einstein tensor
and covariant differentiation corresponding to the metric $g_{\mu
\nu }$ respectively. Solving the field equations
(\ref{fil1})-(\ref{fil3}) directly is a non-trivial task because
the right hand side of (\ref{fil1}) includes the second
derivatives of the scalar. We can remove this difficulty by the
conformal transformation

\begin{eqnarray}
\bar{g}_{\mu \nu } &=&\Phi ^{2/(n-1)}g_{\mu \nu },  \nonumber \\
\bar{\Phi} &=&\frac{n-3}{4\alpha }\ln \Phi ,  \label{conf}
\end{eqnarray}
where
\begin{equation}
\alpha =(n-3)/\sqrt{4(n-1)\omega +4n}  \label{alpha}
\end{equation}
One may note that $\alpha$ goes to zero as $\omega$ goes to
infinity and the
BD theory reduces to Einstein theory. By this transformation, the action (%
\ref{I1}) transforms to

\begin{equation}
\bar{I}_{G}=-\frac{1}{16\pi }\int_{\mathcal{M}}d^{n+1}x\sqrt{-\bar{g}}%
\left\{ \bar{\mathcal{R}}\text{ }-\frac{4}{n-1}(\overline{\nabla }\bar{\Phi}%
)^{2}-\bar{V}(\bar{\Phi})-\exp \left( -\frac{4\alpha \bar{\Phi}}{(n-1)}%
\right) \bar{F}_{\mu \nu }\bar{F}^{\mu \nu }\right\} ,  \label{I2}
\end{equation}
where $\bar{\mathcal{R}}$ and $\ \bar{\nabla}$ are the Ricci
scalar and covariant differentiation corresponding to the metric
$\bar{g}_{\mu \nu }$, and $\bar{V}(\bar{\Phi})$ is
\[
\bar{V}(\bar{\Phi})=\Phi ^{-(n+1)/(n-1)}V(\Phi )
\]
Varying the action (\ref{I2}) with respect to $\bar{g}_{\mu \nu }$, $\bar{%
\Phi}$ and $\bar{F}_{\mu \nu }$, we obtain equations of motion as

\begin{eqnarray}
&& \bar{\mathcal{R}}_{\mu \nu }=\frac{4}{n-1}\left( \bar{\nabla}_{\mu }\bar{%
\Phi}\bar{\nabla}_{\nu }\bar{\Phi}+\frac{1}{4}\bar{V}\bar{g}_{\mu
\nu
}\right) +2e^{-4\alpha \bar{\Phi}/(n-1)}\left( \bar{F}_{\mu \lambda }\bar{F}%
_{\nu }^{\text{ }\lambda }-\frac{1}{2(n-1)}\bar{F}_{\rho \sigma }\bar{F}%
^{\rho \sigma }\bar{g}_{\mu \nu }\right) ,  \label{fild1} \\
&& \bar{\nabla}^{2}\bar{\Phi}=\frac{n-1}{8}\frac{\partial
\bar{V}}{\partial \bar{\Phi}}-\frac{\alpha }{2}e^{-4\alpha
\bar{\Phi}/(n-1)}\bar{F}_{\rho
\sigma }\bar{F}^{\rho \sigma },  \label{fild2} \\
&& \partial _{\mu }\left[ \sqrt{-\bar{g}}e^{-4\alpha \bar{\Phi}/(n-1)}\bar{F}%
^{\mu \nu }\right] =0  \label{fild3}
\end{eqnarray}
Therefore, if $(\bar{g}_{\mu \nu },\bar{F}_{\mu \nu },\bar{\Phi})$
is the
solution of Eqs. (\ref{fild1})-(\ref{fild3}) with potential $\bar{V}(\bar{%
\Phi})$, then

\begin{equation}
\left[ g_{\mu \nu },F_{\mu \nu },\Phi \right] =\left[ \exp \left( -\frac{%
8\alpha \bar{\Phi}}{\left( n-1\right) (n-3)}\right) \bar{g}_{\mu \nu },\bar{F%
}_{\mu \nu },\exp \left( \frac{4\alpha \bar{\Phi}}{n-3}\right)
\right] \label{trans}
\end{equation}
is the solution of Eqs. (\ref{fil1})-(\ref{fil3}) with potential
$V(\Phi )$.

\section{Charged Rotating Solutions In $n+1$ dimensions with $k$ Rotation
Parameters\label{Sol}}

Here we construct the $(n+1)$-dimensional\ solutions of BD theory with $%
n\geq 4$ and the quadratic potential
\[
V(\Phi )=2\Lambda \Phi ^{2}
\]
Applying the conformal transformation (\ref{conf}), the potential $\bar{V}(%
\bar{\Phi})$ becomes
\begin{equation}
\bar{V}(\bar{\Phi})=2\Lambda \exp \left( \frac{4\alpha \bar{\Phi}}{n-1}%
\right) ,  \label{Liovpot}
\end{equation}
which is a Liouville-type potential. Thus, the problem of solving
Eqs. (\ref {fil1})-(\ref{fil3}) with quadratic potential reduces
to the problem of solving Eqs. (\ref{fild1})-(\ref{fild3}) with
Liouville-type potential.

The rotation group in $n+1$ dimensions is $SO(n)$ and therefore
the number of independent rotation parameters for a localized
object is equal to the number of Casimir operators, which is
$[n/2]\equiv k$, where $[n/2]$ is the
integer part of $n/2$. The solutions of the field equations (\ref{fild1})-(%
\ref{fild3}) with $k$ rotation parameter $a_{i}$, and
Liouville-type potential (\ref{Liovpot}) is \cite{SDRP}

\begin{eqnarray}
d\bar{s}^{2} &=&-f(r)\left( \Xi dt-{{\sum_{i=1}^{k}}}a_{i}d\varphi
_{i}\right)
^{2}+\frac{r^{2}}{l^{4}}R^{2}(r){{\sum_{i=1}^{k}}}\left(
a_{i}dt-\Xi l^{2}d\varphi _{i}\right) ^{2}  \nonumber \\
&&-\frac{r^{2}}{l^{2}}R^{2}(r){\sum_{i=1}^{k}}(a_{i}d\varphi
_{j}-a_{j}d\varphi _{i})^{2}+\frac{dr^{2}}{f(r)}+\frac{r^{2}}{l^{2}}%
R^{2}(r)dX^{2},  \nonumber \\
\Xi ^{2} &=&1+\sum_{i=1}^{k}\frac{a_{i}^{2}}{l^{2}},  \nonumber \\
\bar{F}_{tr} &=&\frac{q\Xi }{(rR)^{n-1}} \exp\left(\frac{4\alpha \bar{\Phi}}{%
n-1}\right)  \nonumber \\
\bar{F}_{\varphi r} &=&-\frac{a_{i}}{\Xi }\bar{F}_{tr}.
\label{dil-metric}
\end{eqnarray}
where $dX^{2}$ is the Euclidean metric on $(n-k-1)$-dimensional
submanifold with volume $\omega _{n-k-1}$. Here $f(r)$, $R(r)$ and
$\bar{\Phi}(r)$ are

\begin{eqnarray}
f(r) &=&\frac{2\Lambda (\alpha ^{2}+1)^{2}c^{2\gamma }}{(n-1)(\alpha ^{2}-n)}%
r^{2(1-\gamma )}-\frac{m}{r^{(n-2)}}r^{(n-1)\gamma
}+\frac{2q^{2}(\alpha ^{2}+1)^{2}c^{-2(n-2)\gamma }}{(n-1)(\alpha
^{2}+n-2)r^{2(n-2)(1-\gamma )}},
\label{Fr} \\
R(r) &=&\exp (\frac{2\alpha \bar{\Phi}}{n-1})=\left(
\frac{c}{r}\right)
^{\gamma },  \label{Rr} \\
\bar{\Phi}(r) &=&\frac{(n-1)\alpha }{2(1+\alpha ^{2})}\ln
(\frac{c}{r}), \label{Phir}
\end{eqnarray}
where $c$ is an arbitrary constant and $\gamma =\alpha ^{2}/(\alpha ^{2}+1)$%
. Using the conformal transformation (\ref{trans}), the
($n+1$)-dimensional rotating solutions of BD theory with $k$
rotation parameters can be obtained as
\begin{eqnarray}
ds^{2} &=&-U(r)\left( \Xi dt-{{\sum_{i=1}^{k}}}a_{i}d\varphi
_{i}\right)
^{2}+\frac{r^{2}}{l^{4}}H^{2}(r){{\sum_{i=1}^{k}}}\left(
a_{i}dt-\Xi
l^{2}d\varphi _{i}\right) ^{2}  \label{Met3} \\
&&-\frac{r^{2}}{l^{2}}H^{2}(r){\sum_{i=1}^{k}}(a_{i}d\varphi
_{j}-a_{j}d\varphi _{i})^{2}+\frac{dr^{2}}{V(r)}+\frac{r^{2}}{l^{2}}%
H^{2}(r)dX^{2},  \nonumber
\end{eqnarray}
where $U(r)$, $V(r)$, $H(r)$ and $\Phi (r)$ are
\begin{eqnarray}
U(r) &=&\frac{2\Lambda (\alpha ^{2}+1)^{2}c^{2\gamma (\frac{n-5}{n-3})}}{%
(n-1)(\alpha ^{2}-n)}r^{2(1-\frac{\gamma \left( n-5\right) }{n-3})}-\frac{%
mc^{(\frac{-4\gamma }{n-3})}}{r^{(n-2)}}r^{\gamma
(n-1+\frac{4}{n-3})}
\nonumber \\
&&+\frac{2q^{2}(\alpha ^{2}+1)^{2}c^{-2\gamma (n-2+\frac{2}{n-3})}}{%
(n-1)(\alpha ^{2}+n-2)r^{2[(n-2)(1-\gamma )-\frac{2\gamma
}{n-3}]}},
\label{Ur} \\
V(r) &=&\frac{2\Lambda (\alpha ^{2}+1)^{2}c^{2\gamma (\frac{n-2}{n-3})}}{%
(n-1)(\alpha ^{2}-n)}r^{2(1-\frac{\gamma \left( n-1\right) }{n-3})}-\frac{%
mc^{(\frac{4\gamma }{n-3})}}{r^{(n-2)}}r^{\gamma
(n-1-\frac{4}{n-3})}
\nonumber \\
&&+\frac{2q^{2}(\alpha ^{2}+1)^{2}c^{-2\gamma (n-2-\frac{2}{n-3})}}{%
(n-1)(\alpha ^{2}+n-2)r^{2[(n-2)(1-\gamma )+\frac{2\gamma
}{n-3}]}},
\label{Vr} \\
H(r) &=&(\frac{c}{r})^{\frac{(n-5)\gamma }{n-3}},  \label{Hr} \\
\Phi (r) &=&(\frac{c}{r})^{\frac{2(n-1)\gamma }{n-3}}.  \label{Pr}
\end{eqnarray}
The electromagnetic field becomes:
\begin{eqnarray}
F_{tr} &=&\frac{qc^{(3-n)\gamma }}{r^{(n-3)(1-\gamma )+2}}  \label{Ftr} \\
F_{\varphi r} &=&-\frac{a_{i}}{\Xi }F_{tr}.  \nonumber
\end{eqnarray}
It is worth to note that the scalar field $\Phi (r)$ and
electromagnetic field $F_{\mu \nu }$ become zero as $r$ goes to
infinity. These solutions reduce to the charged rotating solutions
of Einstein gravity as $\omega $ goes to infinity ($\alpha $
vanishes) \cite{Lem,Deh3}. It is also notable to mention that
these solutions are valid for all values of $\omega $.

\subsection{Properties of the solutions}

One can show that the Kretschmann scalar $R_{\mu \nu \lambda
\kappa }R^{\mu \nu \lambda \kappa }$ diverges at $r=0$, and
therefore there is a curvature singularity located at $r=0$.
Seeking possible black hole solutions, we turn to look for the
existence of horizons. As in the case of rotating black hole
solutions of the Einstein gravity, the above metric given by (\ref{Met3})-(%
\ref{Pr}) has both Killing and event horizons. The Killing horizon
is a null surface whose null generators are tangent to a Killing
field. It is easy to see that the Killing vector
\begin{equation}
\chi =\partial _{t}+{{{\sum_{i=1}^{k}}}}\Omega _{i}\partial _{\phi
_{i}}, \label{Kil}
\end{equation}
is the null generator of the event horizon, where $k$ denotes the
number of rotation parameters. Setting $a_{i}\rightarrow ia_{i}$
yields the Euclidean section of (\ref{Met3}), whose regularity at
$r=r_{+}$ requires that we
should identify $\phi _{i}\sim \phi _{i}+\beta _{+}\Omega _{i}$, where $%
\Omega _{i}$'s are the angular velocities of the outer event
horizon. One obtains:
\begin{equation}
\Omega _{i}=\frac{a_{i}}{\Xi l^{2}}.  \label{angvel}
\end{equation}
The temperature may be obtained through the use of definition of
surface gravity,
\begin{equation}
T_{+}=\frac{1}{2\pi }\sqrt{-\frac{1}{2}\left( \nabla _{\mu }\chi
_{\nu }\right) \left( \nabla ^{\mu }\chi ^{\nu }\right) ,}
\end{equation}
where $\chi $ is the Killing vector (\ref{Kil}). One obtains
\begin{eqnarray}
T_{+} &=&\frac{f^{\text{ }^{\prime }}(r_{+})}{4\pi \Xi }=\frac{1}{4\pi \Xi }%
\left( \frac{(n-\alpha ^{2})m}{\alpha ^{2}+1}{r_{+}}^{(n-1)(\gamma -1)}-%
\frac{4q^{2}(\alpha ^{2}+1)b^{-2(n-2)\gamma }}{(\alpha ^{2}+n-2){r_{+}}%
^{\gamma }}{r_{+}}^{(2n-3)(\gamma -1)}\right)  \nonumber \\
&=&-\frac{2(1+\alpha ^{2})}{4\pi \Xi (n-1)}\left( \Lambda
b^{2\gamma
}r_{+}^{1-2\gamma }+\frac{q^{2}b^{-2(n-2)\gamma }}{{r_{+}}^{\gamma }}{r_{+}}%
^{(2n-3)(\gamma -1)}\right) ,  \label{Tem}
\end{eqnarray}
which shows that the temperature of the solution is invariant
under the conformal transformation (\ref{conf}). This result
conclude from this point that the conformal parameter is regular
at the horizon.

As one can see from Eq. (\ref{Ur}), the solution is ill-defined for $%
\alpha^2=n$ with a quadratic potential ($\Lambda \neq 0$). The cases with $%
\alpha^2 >n$ and $\alpha^2 <n$ should be considered separately. In
the first
case where $\alpha^2>n$, as $r$ goes to infinity the dominant term in Eq. (%
\ref{Ur}) is the second term, and therefore the spacetime has a
cosmological horizon for positive values of the mass parameter,
despite the sign of the cosmological constant $\Lambda $. In the
second case where $\alpha^2 <n$, as $r$ goes to infinity the
dominant term is the first term, and therefore there exist a
cosmological horizon for $\Lambda >0$, while there is no
cosmological horizons if $\Lambda <0$ . Indeed, in the latter case ($%
\alpha^2 <n$ and $\Lambda <0$) the spacetimes associated with the solution (%
\ref{Ur})-(\ref{Pr}) exhibit a variety of possible causal
structures
depending on the values of the metric parameters $\alpha $, $m$, $q$, and $%
\Lambda $. One can obtain the causal structure by finding the roots of $%
V(r)=0$. Unfortunately, because of the nature of the exponents in (\ref{Vr}%
), it is not possible to find explicitly the location of horizons
for an arbitrary value of $\alpha $ ($\omega$). But, we can obtain
some information by considering the temperature of the horizons.

Equation (\ref{Tem}) shows that the temperature is negative for
the two
cases of (\emph{i}) $\alpha >\sqrt{n}$ despite the sign of $\Lambda $, and (%
\emph{ii}) positive $\Lambda $ despite the value of $\alpha $. As
we argued above in these two cases we encounter with cosmological
horizons, and therefore the cosmological horizons have negative
temperature. Numerical calculations show that the temperature of
the event horizon goes to zero as the black brane approaches the
extreme case. Thus, one can see from Eq. (\ref {Tem}) that there
exists an extreme black brane only for negative $\Lambda $ and
$\alpha <\sqrt{n}$, if
\begin{equation}
r_{\mathrm{ext}}^{(3-n)\gamma +n-2}=\frac{4q^{2}(1+\alpha
^{2})^{2}b^{-2(n-2)\gamma }}{m_{\mathrm{ext}}(n-\alpha ^{2})(\alpha ^{2}+n-2)%
}
\end{equation}
where $m_{\mathrm{ext}}$ is the extremal mass parameter of black
brane. If one substitutes this $r_{\mathrm{ext}}$ into the
equation $f(r_{ext})=0$, then one obtains the condition for
extreme black brane as:
\begin{equation}
m_{\mathrm{ext}}=\frac{4q^{2}(1+\alpha ^{2})^{2}b^{-2(n-2)\gamma }}{%
(n-\alpha ^{2})(\alpha ^{2}+n-2)}\left( \frac{-\Lambda b^{2\gamma (n-1)}}{%
q_{ext}^{2}}\right) ^{\frac{(3-n)\gamma +n-2}{2(\gamma -1)(1-n)}}
\end{equation}
Indeed, the metric of Eqs. (\ref{Met3})-(\ref{Pr}) has two inner
and outer horizons located at $r_{-}$ and $r_{+}$, provided the
mass parameter $m $ is
greater than $m_{\mathrm{ext}}$, an extreme black brane in the case of $m=m_{%
\mathrm{ext}}$, and a naked singularity if $m<m_{\mathrm{ext}}$.
Note that in the absence of scalar field ($\alpha =\gamma =0$)
$m_{\mathrm{ext}}$ reduces to that obtained in \cite{Deh3}.

Next, we calculate the electric charge of the solutions. To
determine the electric field we should consider the projections of
the electromagnetic field tensor on special hypersurfaces. The
normal to such hypersurfaces is
\[
u^{0}=\frac{1}{N},\text{ \ }u^{r}=0,\text{ \
}u^{i}=-\frac{V^{i}}{N},
\]
and the electric field is $E^{\mu }=g^{\mu \rho }F_{\rho \nu
}u^{\nu }$, where $N$ and $V^{i}$ are the lapse and shift
function. Denoting the volume of the hypersurface boundary at
constant $t$ and $r$ by $V_{n-1}=(2\pi )^{k}\omega _{n-k-1}$, the
electric charge per unit volume $V_{n-1}$ can be found by
calculating the flux of the electric field at infinity, yielding
\[
{Q}=\frac{\Xi q}{4\pi l^{n-2}}
\]
Comparing the above charge with the charge of black brane
solutions of Einstein-Maxwell-dilaton gravity obtained in
\cite{SDRP}, one finds that charge is invariant under the
conformal transformation (\ref{conf}). The electric potential $U$,
measured at infinity with respect to the horizon, is defined by
\cite{Cal}
\begin{equation}
U=A_{\mu }\chi ^{\mu }\left| _{r\rightarrow \infty }-A_{\mu }\chi
^{\mu }\right| _{r=r_{+}} ,  \label{Ch}
\end{equation}
where $\chi $ is the null generators of the event horizon. One can
easily show that the vector potential $A_{\mu }$ corresponding to
electromagnetic tensor (\ref{Ftr}) can be written as
\begin{equation}
A_{\mu }=\frac{qc^{(3-n)\gamma }}{\Gamma r^{\Gamma }}\left( \Xi
\delta _{\mu }^{t}-a_{i}\delta _{\mu }^{i}\right)
\hspace{0.5cm}{\text{(no sum on i)}}, \label{Pot}
\end{equation}
where $\Gamma =\gamma (3-n)+n-2$. Therefore the electric potential
is
\begin{equation}
U=\frac{qc^{(3-n)\gamma }}{\Xi \Gamma {r_{+}}^{\Gamma }}
\label{U}
\end{equation}

\section{Action and Conserved Quantities\label{Therm}}

The action (\ref{I1}) does not have a well-defined variational
principle, since one encounters a total derivative that produces a
surface integral involving the derivative of $\delta g_{\mu \nu }$
normal to the boundary. These normal derivative terms do not
vanish by themselves, but are canceled by the variation of the
surface term
\begin{equation}
I_{b}=-\frac{1}{8\pi }\int_{\partial
\mathcal{M}}d^{n}x\sqrt{-\gamma }K\Phi \label{Ib1}
\end{equation}
where $\gamma $ and $K$ are the determinant of the induced metric
and the
trace of extrinsic curvature of boundary. In general the action $I_{G}+I_{b}$%
, is divergent when evaluated on the solutions, as is the
Hamiltonian and other associated conserved quantities. Rather than
eliminating these divergences by incorporating reference term, a
counterterm $I_{\mathrm{ct}}$ may be added to the action which is
functional only of the boundary curvature invariants. For
asymptotically (A)dS solutions of Einstein gravity, the way that
one deals with these divergences is through the use of counterterm
method inspired by (A)dS/CFT correspondence \cite{Malda}.
However, in the presence of a non-trivial BD scalar field with potential $%
V(\Phi )=2\Lambda \Phi ^{2}$, the spacetime may not behave as either dS ($%
\Lambda >0$) or AdS ($\Lambda <0$). In fact, it has been shown
that with the exception of a pure cosmological constant potential,
where $\alpha =0$, no AdS or dS static spherically symmetric
solution exist for Liouville-type potential \cite{Pollet}. But, as
in the case of asymptotically AdS spacetimes, according to the
domain-wall/QFT (quantum field theory) correspondence \cite{Boon},
there may be a suitable counterterm for the action which removes
the divergences. In this paper, we deal with the spacetimes with
zero curvature boundary, and therefore all the counterterm
containing the curvature invariants of the boundary are zero.
Thus, the counterterm reduces to a volume term as
\begin{equation}
I_{\mathrm{ct}}=-\frac{(n-1)}{8\pi l_{\mathrm{eff}}}\int_{\partial \mathcal{M%
}}d^{n}x\sqrt{-\gamma },\   \label{Ict}
\end{equation}
where $l_{\mathrm{eff}}$ is given by

\begin{equation}
l_{\mathrm{eff}}^{2}=\frac{(n-1)(\alpha ^{2}-n)}{2\Lambda \Phi
^{3/2}} \label{Leff}
\end{equation}
As $\alpha $ goes to zero, the effective $l_{\mathrm{eff}}^{2}$ of
Eq. (\ref {Leff}) reduces to $l^{2}=-n(n-1)/2\Lambda $ of the
(A)dS spacetimes. One may note that the counterterm has the same
form as in the case of asymptotically AdS solutions with zero
curvature boundary, where $l$ is replaced by $l_{\mathrm{eff}}$.
The total action, $I$, can be written as
\begin{equation}
I=I_{G}+I_{b}+I_{\mathrm{ct}}.  \label{Itot}
\end{equation}
The Euclidean actions per unit volume $V_{n-1}$ can then be
obtained through the use of Eqs. (\ref{I1}), (\ref{Ib1}) and
(\ref{Ict}) as
\begin{equation}
I=\beta \frac{(\alpha ^{4}-1)}{16\pi l^{n}}\left( -\frac{nc^{(n+1)\gamma }}{%
\alpha ^{2}-n}r_{+}^{n-(n+1)\gamma }+\frac{2q^{2}l^{2}c^{(3-n)\gamma }}{%
(\alpha ^{2}+n-2)(n-1)}r_{+}^{(n-3)\gamma -n+2}\right)
\end{equation}
It is a matter of calculation to obtain the action as a function
of the intensive quantities $\beta $, $\mathbf{\Omega }$ and $U$
by using the expression for the temperature, the angular velocity
and the potential given in Eqs. (\ref{angvel}), (\ref{Tem}) and
(\ref{U}) as
\begin{equation}
I=\frac{\beta (\alpha ^{4}-1)c^{(n-3)\gamma }}{16\pi l^{n}}\left( -\frac{%
nc^{4\gamma }r_{+}^{n-(n+1)\gamma }}{\alpha
^{2}-n}-\frac{2U^{2}(\alpha ^{2}+n-2)l^{2}r_{+}^{(n-2)(1-\gamma
)+\gamma }}{(n-1)(\Omega ^{2}l^{2}-1)(\alpha ^{2}+1)^{2}}\right) ,
\label{Smar}
\end{equation}
where $r_{+}$ is

\begin{equation}
r_{+}=\left[ -\frac{\left( \pi (n-1)+\sqrt{\left[ \pi
^{2}(n-1)^{2}-\Lambda U^{2}\beta ^{2}(\alpha ^{2}+n-2)^{2}\right]
}\right) }{\beta \Lambda
(1+\alpha ^{2})\sqrt{1-\Omega ^{2}l^{2}}}c^{-2\gamma }\right] ^{\frac{1}{%
1-2\gamma }}
\end{equation}

Since the Euclidean action is related to the free energy in the
grand-canonical ensemble, the electric charge $Q$, the angular
momentum $J_{i}$, the entropy $S$ and the mass $M$ can be found
using the familiar thermodynamics relations:
\begin{eqnarray*}
Q &=&-\beta ^{-1}\frac{\partial I}{\partial U}=\frac{\Xi q}{4\pi l^{n-2}}, \\
J_{i} &=&-\beta ^{-1}\frac{\partial I}{\partial \Omega _{i}}=\frac{%
c^{(n-1)\gamma }}{16\pi l^{n-2}}\left( \frac{n-\alpha ^{2}}{1+\alpha ^{2}}%
\right) \Xi ma_{i}, \\
S &=&\left( \beta \frac{\partial }{\partial \beta }-1\right)
I=\frac{\Xi c^{(n-1)\gamma }}{4l^{\left( n-2\right)
}}r_{+}^{(n-1)\left( 1-\gamma
\right) }, \\
M &=&\left( \frac{\partial }{\partial \beta }-\beta ^{-1}U\frac{\partial }{%
\partial U}-\beta ^{-1}\sum \Omega _{i}\frac{\partial }{\partial \Omega _{i}}%
\right) I=\frac{c^{(n-1)\gamma }}{16\pi l^{n-2}}\left(
\frac{(n-\alpha ^{2})\Xi ^{2}+\alpha ^{2}-1}{1+\alpha ^{2}}\right)
m
\end{eqnarray*}
For $a_{i}=0$ ($\Xi =1$), the angular momentum per unit length
vanishes, and therefore $a_{i}$ is the $i$th rotational parameter
of the spacetime. One may note that the charge $Q$ calculated
above coincides with Eq. (\ref{Ch}) It is worth to note that the
area law is no longer valid in Brans-Dicke theory
\cite{Zan1,fail}. Nevertheless, the entropy remains unchanged
under conformal transformations. Comparing the conserved and
thermodynamic
quantities calculated in this section with those obtained in Ref. \cite{SDRP}%
, one finds that they are invariant under the conformal
transformation (\ref {trans}). Straightforward calculations show
that these quantities calculated satisfy the first law of
thermodynamics,

\begin{equation}
dM=TdS+{{{\sum_{i=1}^{k}}}}\Omega _{i}dJ_{i}+Ud{Q}
\label{First-law}
\end{equation}

\section{Closing Remarks}

Till now, no charged rotating black hole solutions has been
constructed for an arbitrary value of coupling constant $\omega$.
In this paper, we presented a class of exact charged rotating
black brane solutions in Brans-Dicke theory with a quadratic
scalar field potential for an arbitrary value of $\omega$ and
investigated their properties. We found that these solutions are
neither asymptotically flat nor (A)dS. These solutions which
exist only for $\alpha^2 \neq n$ have a cosmological horizon for (\emph{i}) $%
\alpha^2 >n$ despite the sign of $\Lambda $, and (\emph{ii})
positive values of $\Lambda $, despite the magnitude of $\alpha $.
For $\alpha^2 <n$, the solutions present black branes with outer
and inner horizons if $m>m_{ext}$,
an extreme black hole if $m=m_{ext}$, and a naked singularity if $m<m_{ext}$%
. The Hawking temperature is negative for inner and cosmological
horizons, and it is positive for outer horizons. We computed the
finite action through the use of counterterm method and obtained
the thermodynamic and conserved quantities of the solutions by
using the relation between the action and free energy. We found
that the entropy does not follow the area law. We also found that
the conserved and thermodynamic quantities are invariant under the
conformal transformation (\ref{trans}) and satisfy the first law
of thermodynamics.
\acknowledgments{This work has been supported
by Research Institute for Astronomy and Astrophysics of Maragha,
Iran}

\end{document}